\documentclass[aps,prl,amsmath,twocolumn,amssymb,floatfixng,showpacs,
superscriptaddress,footinbib]{revtex4-1}
\pdfoutput=1
\usepackage[dvips]{graphics}
\usepackage{bm}
\usepackage{float}
\usepackage{epsfig}
\usepackage{enumerate}
\usepackage{amsmath}
\usepackage{color}
\usepackage{graphicx}
\usepackage[colorlinks=true,linktoc=page,linkcolor=red,citecolor=blue]{hyperref}

\newcommand\bea{\begin{eqnarray}}
\newcommand\eea{\end{eqnarray}}
\newcommand\beq{\begin{equation}}  
\newcommand\eeq{\end{equation}}

\begin{document}

\title{\titlename}
\date{\today}
\title{Tailoring higher-order van Hove singularities in non-Hermitian interface systems via Floquet engineering}

\author{Ayan Banerjee}
\thanks{These two authors contributed equally}
\email{ayanbanerjee@iisc.ac.in}
\affiliation{Solid State and Structural Chemistry Unit, Indian Institute of Science, Bangalore 560012, India}
\author{Debashree Chowdhury}
\thanks{These two authors contributed equally}
\email{debashreephys@gmail.com}
\affiliation{Centre of Nanotechnology, Indian Institute of Technology Roorkee, Roorkee, Uttarakhand-247667}
\author{Awadhesh Narayan}
\email{awadhesh@iisc.ac.in}
\affiliation{Solid State and Structural Chemistry Unit, Indian Institute of Science, Bangalore 560012, India}

\date{\today}

\begin{abstract}
We propose a non-Hermitian (NH) interface system formed between two NH nodal line semimetals driven by optical fields as a platform for generation and tailoring of higher-order van Hove singularities (VHS). Through an analytical analysis of the density of states (DOS), we find VHS with logarithmic divergences in the Hermitian limit. Upon introducing NH terms, four exceptional rings on two sides of the Fermi line are formed. By tuning the NH parameters and the light amplitude, we find a situation when one exceptional ring crosses the Fermi line, where a saddle point appears and results in a paired VHS around the origin. In contrast, when an exceptional contour resides at the Fermi energy, the saddle points critically get destroyed and we obtain a single peak in the DOS, with power-law divergences. These higher-order divergences that appear in an NH system have a different origin than that of the higher-order VHS in Hermitian systems, where no saddle point merging is noted. Our results suggest NH interfaces to be promising avenues for exploring higher-order VHS.
\end{abstract}

\maketitle

%\section{Introduction}
The advent of topological phases has become an intriguing aspect of condensed matter physics~\cite{hasan2010colloquium,vafek2014dirac,wen2017colloquium}. Starting with the unique features of two- and three-dimensional topological insulators~\cite{kane2005quantum,bernevig2006quantum,hasan2011three}, the new focus now are the semimetallic phases that appear in different systems. What makes these semi-metal phases alluring are the isolated band-crossing points or lines, which may result from band inversion~\cite{burkov2016topological,armitage2018weyl}. These unique band crossings are related to the monopoles of Berry phases. So far, three different kinds of semimetals are among the most explored, namely Dirac semimetals~\cite{chiu2016classification,murakami2007phase,young2012dirac,liu2014discovery}, Weyl semimetals~\cite{armitage2018weyl,yan2017topological}, and nodal line semimetals~\cite{shao2020electronic,fang2016topological,burkov2011topological}. In the case of Dirac semimetals fourfold degenerate Dirac points appear in the system, which are protected by crystalline symmetries. In Weyl semimetals, the band crossing points, coined as Weyl points are doubly degenerate and have definite chiralities~\cite{hasan2017discovery,vanderbilt2018berry}. Furthermore, the topological nodal line semimetals form degeneracy lines in lieu of discrete points in the momentum space~\cite{fang2016topological,chiu2016classification}.

\begin{figure}	
\includegraphics[width=0.9\linewidth]{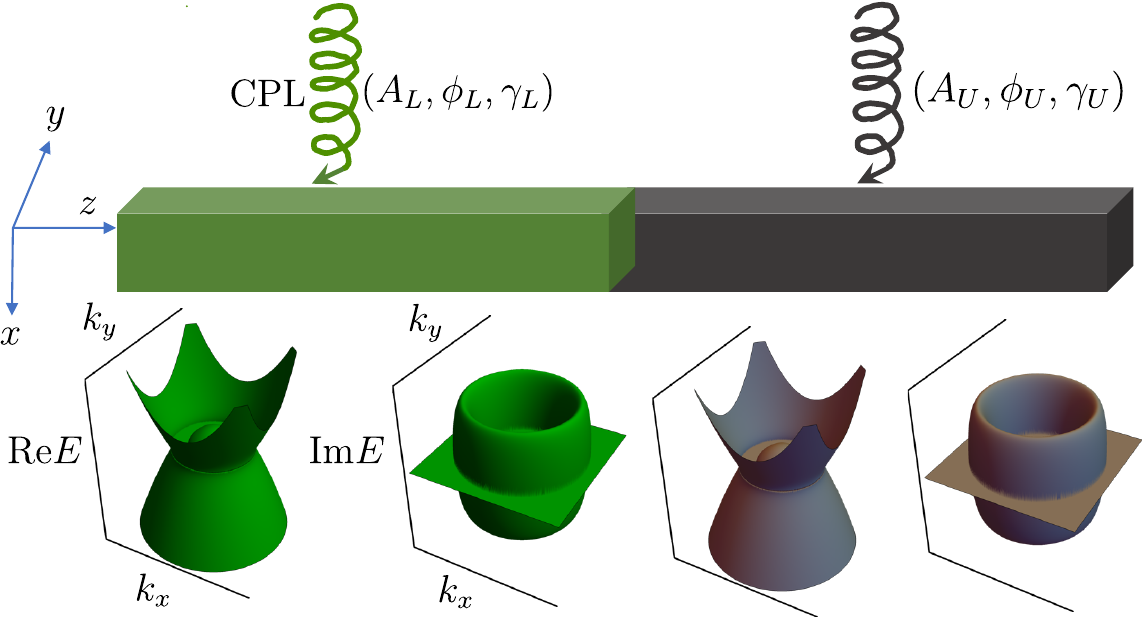}
\caption{\textbf{Conceptual illustration of the NH interface system.} The schematic representation illustrates the arrangement of the non-Hermiticity-induced nodal line semimetals divided into two regions -- lower $(z < 0)$ and upper $(z > 0)$ half-spaces. These regions are subjected to high-frequency monochromatic circularly polarized light (CPL) irradiation with distinct intensities, polarizations, and non-Hermiticity coefficients denoted as $(A_L, \phi_L,\gamma_L)$ for the lower half-space and $(A_U, \phi_U,\gamma_U)$ for the upper half-space. In the presence of non-Hermiticity, the nodal ring splits into two exceptional rings. Driving allows the tuning of exceptional physics.}
\label{schematic}	
\end{figure}

On the other hand, topological NH systems are at the forefront of research for condensed matter, optics, and photonics communities~\cite{bergholtz2021exceptional,ashida2020non,ding2022non,banerjee2023non,lee2016anomalous,lin2023topological,kunst2018biorthogonal,torres2019perspective,borgnia2020non,zhang2022review,yao2018non,zhao2019non,miri2019exceptional}. The distinctive feature of NH systems is the existence of certain degenerate points where both the eigenenergies as well as eigenfunctions of the system coalesce. These degenerate points are known as exceptional points (EP)~\cite{kato2013perturbation}. These EPs endow unique features in NH topological systems~\cite{kawabata2019classification,yang2019non,budich2019symmetry}. Moreover, in NH systems, enhanced tunability can be achieved by illuminating the system following the well-established principles of Floquet engineering~\cite{zhou2023non,banerjee2020controlling}. Hermitian topological systems show tunable Fermi surface topology in the presence of time-periodic fields~\cite{oka2009photovoltaic,cayssol2013floquet,gu2011floquet,perez2014floquet,chen2018floquet,chan2016chiral,yan2016tunable,narayan2016tunable}. In NH topological systems, however, the understanding of the topology caused due to the combined action of non-Hermiticity and driving is a topic of recent thrust~\cite{zhou2023non}. For instance, circular driving generates new exceptional contours and can cause topological charge division~\cite{chowdhury2021light}. 

It is well known that a saddle point in the band structure could cause divergences in the DOS, which is coined as the VHS~\cite{grosso2013solid}. This logarithmic divergence in DOS, which when lying at the Fermi energy, leads to intriguing physics. The VHS spawns effects such as superconductivity, charge, and spin density waves. Interestingly, VHS has intriguing effects on topological systems~\cite{wu2021chern}. Recently, it has been found that deviations from the usual logarithmic singularity may be engineered, resulting in higher-order singularities with a power-law divergence~\cite{yuan2019magic,shtyk2017electrons}. These divergences are the crucial ingredients in understanding the ordering instabilities in systems such as stacked bilayer graphene, twisted bilayer transition-metal dichalcogenides, or even high Tc superconductors and heavy fermions materials~\cite{guerci2022higher,wang2021moire,oriekhov2021orbital,kerelsky2019maximized,markiewicz2021high}. In a recent work, it was shown that high-frequency light can induce a logarithmic-type VHS in an interface Hermitian system~\cite{bonasera2022tunable}, in the presence of asymmetric light intensities. This raises the question of the nature of the DOS and VHS in non-Hermitian interface systems. 

The system we propose contains two nodal line semimetals, with different gain/loss terms and the two halves are differently illuminated with two opposite circular polarized monochromatic light beams. Fig.~\ref{schematic} shows a schematic of our proposal. In this scenario, the energy spectra and the DOS show unique features, that are absent in the Hermitian limit. The band structure of the system is rich enough, resulting in regions where the Fermi line falls within or on one of the exceptional contours, which shows unique Fermi surface topology. When non-Hermiticity is introduced on both sides $(\gamma_L,\gamma_U>0)$, the two nodal rings of the Hermitian limit are broken into four exceptional rings (located on the two sides of the Fermi line $\mathrm{Re}(E)=0$). Consequently, one obtains eight EPs along the $k_x $ axis. When one of such exceptional rings crosses the Fermi line, a saddle point at a high symmetry point $(k_{x}=0,k_{y}=0)$ appears, resulting in a paired VHS around the origin. If an exceptional contour resides on the Fermi energy, the saddle points critically get destroyed along the $k_x$ axis, and a single peak in the DOS is obtained, which at a higher resolution splits into two peaks with the emergence of higher order ($E^{1/3}$ and $E^{1/2}$) power law divergences. In NH systems, these higher-order singularities appear not because of the saddle points but rather due to saddle point coalescing~\cite{hu2022geometric}.

%\section{The model Hamiltonian}
The Hamiltonian of the nodal ring semimetal with the NH term is written as~\cite{wang2019non}

\begin{align}\label{1}
H(k) =\Big[m- B\Big(k_{x}^{2}+k_{y}^{2}+k_{z}^{2}\Big)\Big]\sigma_{x} + i\gamma\sigma_{y} + v_{z}k_{z}\sigma_{z},
\end{align}

where $\sigma_i$ ($i = x, y, z$) are Pauli matrices that represent the two orbitals and $v_z$ is the Fermi velocity in the direction of $k_z$. Additionally, $m$ and $B$ are parameters characterized by energy units and inverse energy units, respectively. In the Hermitian limit, when $mB > 0$, the conduction and valence bands intersect, forming a nodal ring in the $k_z = 0$ plane at $k_x^2 + k_y^2 = m/B$~\cite{yan2016tunable,narayan2016tunable}. Conversely, for $mB < 0$, the system exists within the trivial insulator phase, featuring an energy gap. To simplify matters, and without loss of generality, it is assumed from here on that $m, B,$ and $v_z$ are all positive unless explicitly stated otherwise. This Hermitian nodal ring remains protected by the joint symmetries of inversion ($P$) and time reversal ($T$), represented as $PT$~\cite{yan2016tunable}. In the presence of an NH term $i\gamma$ that accounts for the dissipative coupling manifesting ``imaginary Zeeman fields"~\cite{lee1952statistical,yao2018non}, the original nodal ring breaks into two exceptional rings along $k_z=0$. The NH term inherently breaks the $PT$ symmetry but respects the chiral symmetry. As $\gamma$ increases, the inner exceptional ring progressively contracts and eventually disappears beyond a critical point at $\gamma = m$, condensing into a point.

\begin{figure*}
 \includegraphics[scale=0.45]{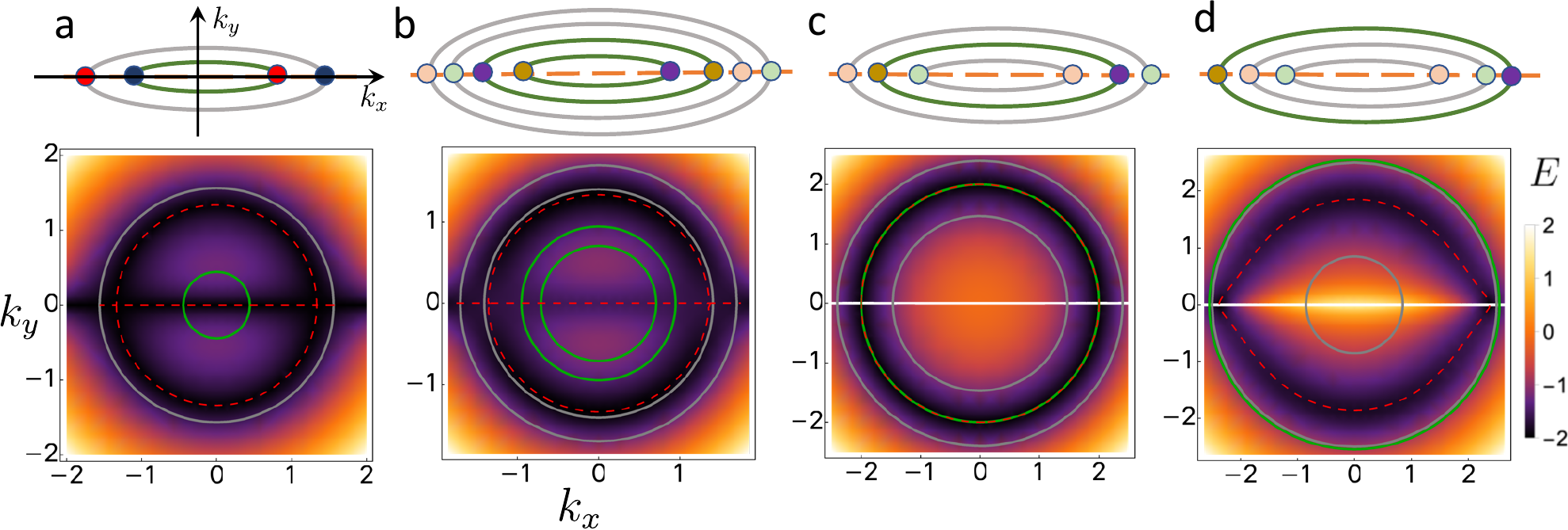}
\caption{\textbf{Illustration of the nodal exceptional rings delineating the Fermi surface spectral topology and energy dispersion of the interface states.} The upper panel shows the (a) nodal ring semimetals in the Hermitian limit $(\gamma_U,\gamma_L=0)$, (b)-(d) exceptional rings (shown in grey and green) for the interface system as a function of light intensity. The two nodal rings arising from the interface system under NH perturbation split into four exceptional rings. Consequently, the four Weyl points with opposite chiralities (red and blue dots) along the $k_x$ split into eight EPs. These four exceptional rings consisting of eight EPs move along the $k_x$ axis and can annihilate each other as a function of non-Hermiticity and driving amplitude. The lower panel depicts the density plots of the interface states. The dotted red lines represent the Fermi line $\text{Re}E(k_{\perp})=0$, and we set $v_z = 1$, $\omega =10$, $B=0.5$, $m=2$, $\phi_U=\pi$ and $\phi_L=\pi$. (a) The interface states are bounded by nodal rings residing on the two sides of the Fermi line. The nodal lines define the coalescing of the interface band into a conduction (valence) bulk band for $A_U=1.25$ and $A_L=1.95$. (b) The nodal lines split into four exceptional rings (on the two sides of the Fermi line), which delimit the interface states for $(A_U,\gamma_U)=(1.25,0.23)$ and $(A_L,\gamma_L)=(1.95,0.35)$. (c) The critical case with six EPs where the green exceptional ring merges on the Fermi line and the grey ring crosses the Fermi line for $(A_U,\gamma_U)=(0.2,0.9)$ and $(A_L,\gamma_L)=(0.0,0.0)$. (d) The green ring traverses the Fermi line, and two grey rings are located on the two sides of the Fermi line for $(A_U,\gamma_U)=(0.72,1.38)$ and $(A_L,\gamma_L)=(1.18,1.94)$.} \label{dis-1}
\end{figure*}

For $k_{z}=0$ we have, $\Big(k_{x}^{2}+k_{y}^{2}\Big)=\frac{m\pm \gamma}{B}$, which gives the nodal ring in the $k_x-k_y$ plane. Next, we drive the system with light polarized along the $y-z$ plane and the vector potential associated with it is
$\bm{A}(t)=A_{} \Big( 0, \cos{\left(\omega t\right)},\sin{\left(\omega t +\phi\right)} \Big)$. Here $A_{}$ and $\omega$ are the amplitude and frequency of the incident light. Also, $\phi=0$ ($\phi=\pi$) corresponds to the right (left) circular polarization. Application of periodically time-driven light allows us to use the Floquet formalism, which in the high-frequency limit provides the effective Hamiltonian~\cite{oka2009photovoltaic,cayssol2013floquet,bukov2015universal,holthaus2015floquet,oka2019floquet,rudner2020band,rudner2020floquet,harper2020topology,sen2021analytic,rodriguez2022quantum,mori2023floquet,rahav2003effective,zhou2023non}.

\begin{align}\label{3}
    H_{F}(k)&=\Big[\tilde{m}- B\Big(k_{x}^{2}+k_{y}^{2}\Big)\Big]\sigma_{x}+ v_{z}k_{z}\sigma_{z}\nonumber\\& + \Big(\lambda k_{y}+i\gamma\Big)\sigma_{y} ,
\end{align}

where $\tilde{m}=m-B e^{2}A_{}^{2}$ and $\lambda=-\frac{2e^{2}B v A_{}^{2}\cos\phi}{\omega}.$ The corresponding energy eigenvalues are 

\begin{align}
&E_{\pm}(k)\nonumber\\&=\pm \sqrt{\Big[\tilde{m}- B\Big(k_{x}^{2}+k_{y}^{2}\Big)\Big]^{2}+v_{z}^{2}k_{z}^{2}+\Big(\lambda k_{y}+i\gamma\Big)^{2}}.
\label{light-eps}
\end{align}

The exceptional degeneracies delineating a pair of EPs  for $\lambda\neq 0$ and $k_{y}=k_{z}=0,$ are found at

\begin{align}\label{5}
    k_{x}=\pm \sqrt{\frac{\tilde{m}\pm \gamma}{B}}.
\end{align}

The exceptional degeneracies, which appear for $\text{Re}(E)=0$ and $\text{Im}(E)=0$, are accompanied by a Fermi arc along the $k_x$ axis protected by nontrivial Z topology for a real line gap~\cite{kawabata2019symmetry}. We note that the chirality of the exceptional rings depends simply on the handedness, $\cos{\phi}$, of the incident laser beam. Moreover, the shape and position of the exceptional rings are tunable by changing the direction and amplitude of the incident laser beams. The topological characterization, as well as Hall signatures in this light-driven system, have been recently studied in the literature~\cite{he2020floquet,wu2020floquet}. Here we examine the NH interface system and explore the exceptional VHS arising from Lifshitz transitions with striking changes in the Fermi surface topology, considering the interplay of laser driving and non-Hermiticity. 

We consider an interface of two nodal ring semimetals with different light intensities and NH parameters on the two sides (see Fig.~\ref{schematic}). The effective Floquet Hamiltonian for each side is written as

\begin{align}\label{6}
    H_{j}(k)=\Bigg[\tilde{m}_{j}-B k^{2}\Big]\sigma_{x}+\Big(\lambda_{j} k_{y}+i\gamma_j\Big)\sigma_{y} + v_{z}k_{z}\sigma_{z},
\end{align}

where $\tilde{m}=m-B e^{2}A_{j}^{2}$, $\lambda_{j}=-\frac{2e^{2}B v A_{j}^{2}\cos\phi_{j}}{\omega}$ with $j\in U,L$, and $k^{2}=k_{x}^{2}+k_{y}^{2}+k_{z}^{2}.$ The interface is considered along the $z$ direction and is located at $z=0$. This system is subjected to the influence of two light beams with opposite circular polarizations. When we consider the high-frequency scenario, we employ an effective Floquet Hamiltonian for each of these regions~\cite{oka2019floquet}. Specifically, we label the right $(z > 0)$ and left $(z < 0)$ half-spaces as $H_U$ and $H_L$ respectively. In Fig.~\ref{schematic}, we have presented a schematic of our setup. Here, the subscripts $L$ and $U$ signify the left and right half-spaces, respectively. However, the sign of the product $\lambda_U\lambda_L$ is negative due to the contrasting circular polarizations of the two light beams. Notably, the assumption we make about the interface between these regions having an abrupt sharpness does not diminish the validity of our results, as the emergent topological interface modes are resilient against perturbations at the interface.

The stationary Schr\"{o}dinger equation, linked to each half-space, is given by

\begin{align}
	\mathcal{H}_j \left(\bm{k}_{\perp},k_z \to -i\partial_{z}\right) \psi_j \left(\bm{r}\right) = E\left(\bm{k}_\perp \right) \psi_j \left(\bm{r}\right),
 \label{waveequation}
\end{align}

where $\bm{k}_{\perp}=\left(k_x,k_y\right)$ and $E\left(\bm{k}_\perp \right)$ is the eigenenergy. We consider the following ansatz wavefunction 

\begin{equation}
\psi_{j}(r)= e^{i k_x x}e^{i k_y y}\big(\begin{smallmatrix}
  \psi_1^{j} \\
  \psi_2^{j}
\end{smallmatrix}\big)e^{ \mu_j z}.
\label{wavefunction}
\end{equation}

The real part of $\mu_j z$ $[\text{Re}(\mu_j z) < 0]$ dictates the spatial localization in proximity to the plane $z = 0$. Substituting Eq.~\ref{wavefunction} into Eq.~\ref{waveequation} yields the secular equation governing the eigenstates

\begin{equation}
\text{det}[H(\bm{k}_{\perp}, \partial_z \rightarrow \mu_j ) -E\mathcal{I}]=0,
\end{equation}

where $\mu_j=\frac{1}{v_z}\sqrt{[\tilde{m}_{j}-B k^2]^2+(\lambda k_y+ i \gamma)^2-E^2}$. The interface energy for each half-space is written as~\cite{bonasera2022tunable}

\begin{widetext}
\begin{align}\label{17}
	E\left(\bm{k}_\perp \right)= \frac{\Big(-i\alpha_{L}\gamma_{U}+i\alpha_{U}\gamma_{L}+k_{y}\alpha_{U}\lambda_{L}-k_{y}\alpha_{L}\lambda_{U}\Big)}{\sqrt{\left(\alpha_{L}-\alpha_{U}\right)^2 - \left(\gamma_{U}-\gamma_{L}+ i k_y\Big(\lambda_{L} - \lambda_{U}\Big)^2 \right)}},
\end{align}
where $\alpha_{j}=\tilde{m}_{j}-B k^{2}.$
\end{widetext}

For each perpendicular momentum component $\bm{k}_{\perp}=(k_x,k_y)$, the quantity $E(\bm{k}_{\perp})$ is the eigenenergy associated with an interface state. This interface energy is defined when both ${\rm Re} [\mu_L] $ and ${\rm Re} [\mu_R ]$ are non-zero. Conversely, if either ${\rm Re}[\mu_L] $ or ${\rm Re}[\mu_R ]$ is zero, the obtained solution characterizes a delocalized mode across the left or right half-space corresponding to a bulk state. Consequently, the condition $Re[\mu_{L}\mu_{R}] =0$ serves as a defining criterion for establishing the boundaries within the two-dimensional momentum space of interface states. The topological attributes of the interface states are contingent upon the spatial arrangement in the two-dimensional momentum space $(k_x - k_y)$ of the surface projections of EPs from the upper and lower exceptional nodal ring semi-metals (ENRS). In the present case, this arrangement is dictated by the varying light intensities experienced by the two half-spaces of the ENRS. Precisely, for each half-space, the manipulation of light intensity $\lambda_j$, as facilitated by Eq.~\ref{light-eps}, results in the ability to either bring the EPs closer to or move them further away from the origin of momenta. The bulk Fermi line of this interface is defined as ${\rm Re}[E(k_{\perp})]=0$ (the red dotted line in Fig.~\ref{dis-1}). Fig.~\ref{dis-1} shows the density plots depicting the energy dispersion characteristics of interface states as a function of $k_{\perp}$. In the Hermitian limit $(\gamma_j=0)$, in the presence of laser driving $(A_{U}=1.25; A_{L}=1.95)$, the interface states are demarcated by the nodal lines, which are the solutions of $\mu_L=\mu_R =0$. Notably, along the nodal lines, the interface band coalesces with the bulk conduction and valence bands. Next, switching on the non-Hermiticity enables the splitting of nodal rings to exceptional rings endowing Lifshitz transitions. Consequently, the interface states become bounded by these exceptional rings. Intriguingly, by manipulating the interplay between light intensity and non-Hermiticity, the exceptional rings undergo movement in the momentum space and traverse the Fermi line (indicated by the red dotted line). The analytical conditions for the three qualitatively different non-Hermitian phases corresponding to Fig.~\ref{dis-1} are as follows  
\begin{widetext}
\noindent (i)  
\begin{align}
\sqrt{\frac{{\tilde{m}_L \pm \gamma L}}{B}} < \frac{{\sqrt{A_U^2 (-A_L^2 B + m) \cos(\phi_L) + A_L^2 (A_U^2 B - m) \cos(\phi_U)}}}{\sqrt{B (A_U^2 \cos(\phi_L) - A_L^2 \cos(\phi_U))}} < \sqrt{\frac{{\tilde{m}_U \pm \gamma_U}}{B}},
\end{align} (ii) 
\begin{align}
\sqrt{\frac{{\tilde{m}_U - \gamma_U}}{B}} < \sqrt{\frac{{\tilde{m}_L \pm \gamma_L}}{B}} = \frac{{\sqrt{A_U^2 (-A_L^2 B + m) \cos(\phi_L) + A_L^2 (A_U^2 B - m) \cos(\phi_U)}}}{\sqrt{B (A_U^2 \cos(\phi_L) - A_L^2 \cos(\phi_U))}} < \sqrt{\frac{{\tilde{m}_U + \gamma_U}}{B}},
\end{align}
(iii) 
\begin{align}
\sqrt{\frac{{\tilde{m}_U - \gamma_U}}{B}} < \frac{{\sqrt{A_U^2 (-A_L^2 B + m) \cos(\phi_L) + A_L^2 (A_U^2 B - m) \cos(\phi_U)}}}{\sqrt{B (A_U^2 \cos(\phi_L) - A_L^2 \cos(\phi_U))}} < \sqrt{\frac{{\tilde{m}_U + \gamma_U}}{B}} < \sqrt{\frac{{\tilde{m}_L \pm \gamma_L}}{B}}.
\end{align}
\end{widetext}

These three conditions correspond to Fig.~\ref{dis-1}(b), (c) and (d), respectively. These, in turn, lead to noteworthy implications in the DOS, which we delve into next. Using the interface energy, the spectral function can be obtained as

\begin{align}
{\cal A}(\omega)=-\frac{1}{\pi}{\rm Im}\Big[\omega+i\eta-E\left(\bm{k}_\perp \right)\Big]^{-1}.
\end{align}

The DOS can then be obtained as $\rho(\omega)=\frac{1}{2\pi}{\cal A}(\omega).$ We note that, in NH systems, employing a bi-orthogonal basis composed of both right and left eigenvectors~\cite{brody2013biorthogonal}, one can construct the NH adaptation of Lehmann's representation of the Green's function~\cite{chen2018hall}.  

In Ref.~\cite{bonasera2022tunable}, the authors have discussed the DOS for two situations. For a symmetric interface with equal light amplitude, the DOS shows a $\Theta$ function behaviour. However, for unequal light amplitudes applied in the two regions of the nodal line semimetals, a VHS is found to appear in the DOS at the Fermi energy. The VHS in the DOS manifests from the existence of saddle points in the energy dispersion of systems. In two-dimensional Hermitian systems characterized by an energy dispersion $E(k_x,k_y)$, a VHS featuring a logarithmically diverging DOS arises at a saddle point $\mathbf{k}_s$, which is governed by $\nabla_{\mathbf{k}}E=0$ and $\text{det}D<0$, where $D_{ij}=\frac{1}{2}\partial_i\partial_j E$ is the Hessian matrix of $E$ at $\mathbf{k}_s$. For instance, consider a Taylor expanded energy dispersion around a saddle point $\mathbf{k}_s$, as $E-E_{\xi}=-a k_x^2+b k_y^2$, where $E_{\xi}$ is the VHS energy. The saddle point criteria are satisfied with the condition $ab <0$, where the two coefficients $-a$ and $b$ are the eigenvalues of the Hessian matrix $D$. 

\begin{figure}	
\includegraphics[width=0.95\linewidth]{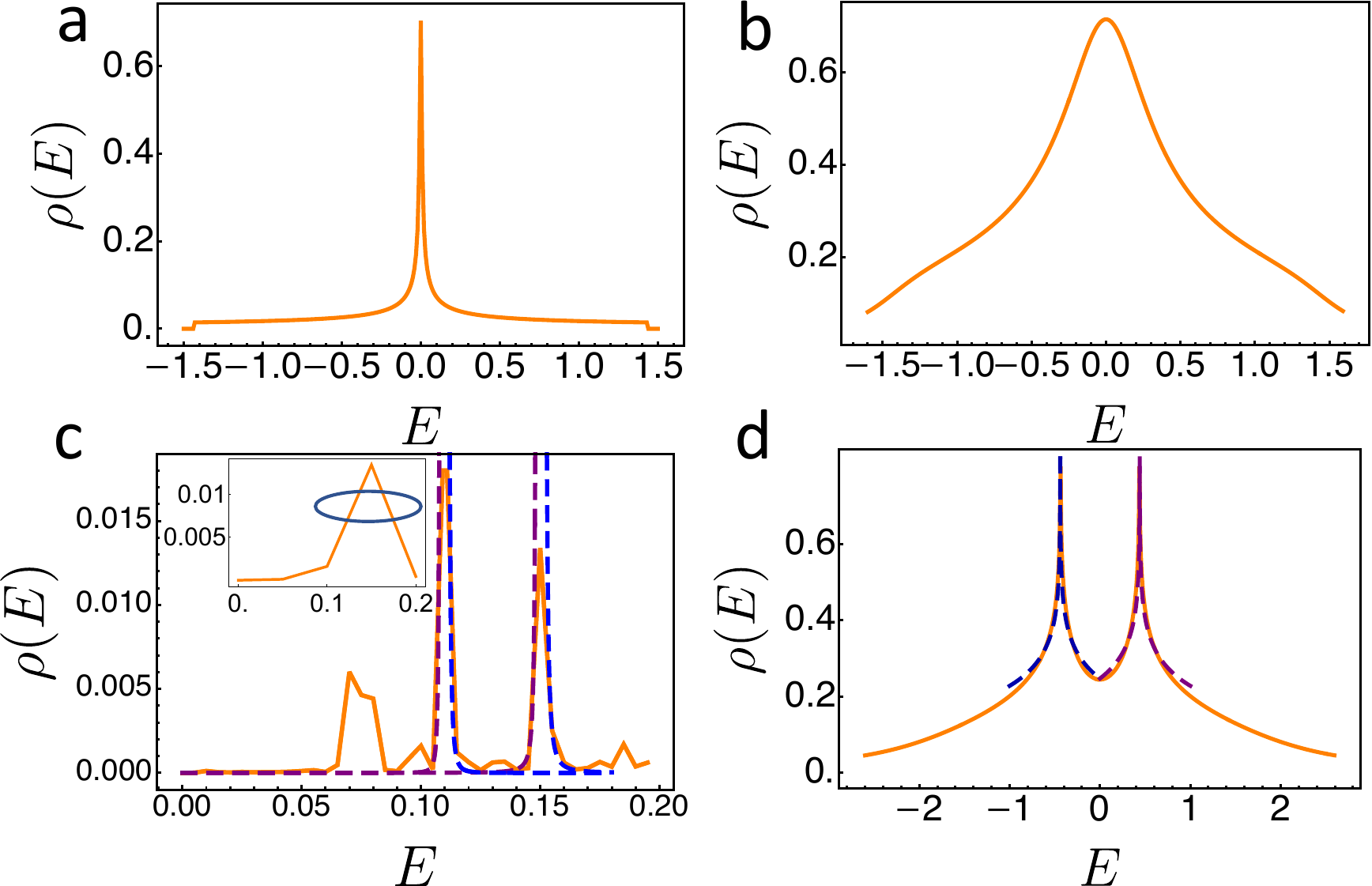}
\caption{\textbf{Density of interface states and VHS physics}. (a) The divergence in the DOS manifests in a VHS for the Hermitian case. (b) Switching on the non-Hermiticity leads to broadening in the DOS as long as the green and grey rings lie on the two sides of the Fermi line. (c) The critical case corresponds to a single peak in the DOS, which gives rise to $n$-th root singularities with $n=2$ and $3$ describing higher order VHS. (d) DOS for paired VHS is symmetric about the origin. The dotted blue lines in (c) and (d) designate the power law fitting with $E^{1/2}$ ($~E^{1/3}$) and $\log(1/|E|)$ logarithm scaling, respectively.}
\label{vanhove}	
\end{figure}

We next analyze the saddle point physics and VHS, in our NH interface system, considering the interplay of exceptional topology and Fermi surface crossing. We inspect the qualitatively different regimes upon tuning of light intensity and non-Hermiticity coefficient, which enable a topological transition of saddle points. First, we consider the Hermitian case $(\gamma_U=\gamma_L=0)$. We obtain two Weyl rings, consisting of four Weyl points along $k_x$ (red and blue points designate the positive and negative chirality; see Fig.~\ref{dis-1}). Since the intricate structure of the VHS around this transition point is solely characterized by local energy dispersion near $\mathbf{k}_s$, we expand $E(\mathbf{k})$ near $k_s$ to higher orders, $E-E_{\xi}=a k_y + b k_y^3 + (d k_y + e k_y^3) k_{x}^2$. The behavior of the VHS depends crucially on the sign of the coefficients $a,b,c,d$ and $e$. For $a/d <0$ two saddle points appear along $k_y=0$ line at momenta $(\pm \frac{i \sqrt{a}}{\sqrt{d}},0)$ with the condition $4ad<0$ resulting in VHS at $E_{\xi}=0$ [see Fig.~\ref{vanhove} (a)]. 

Next, we move on to different NH cases presented in Fig.~\ref{dis-1}. Importantly, we focus on the real part of the expanded dispersion near the saddle points, considering the fact that the imaginary parts add broadening to the spectrum. We expand $E$ near $k_s$ to higher orders, $E-E_{\xi}=\frac{1}{\sqrt{\Theta}}\big(\alpha_i+\beta k_y+ \delta_{i} k^2_y + \Gamma k_y^3 + (\zeta_i+\eta k_y + \epsilon_i k_y^2) k_x^2\big)$. The subscript $``i"$ denotes the purely imaginary coefficients. We switch on the non-Hermiticity $(\gamma_L,\gamma_U>0)$ to obtain four exceptional rings (located on the two sides of the Fermi line $\mathrm{Re}[E]=0$). Consequently, four Weyl points are split into eight EPs along the $k_x$ axis. Interestingly, in this case, the saddle points positions are renormalized on the $k_y=0$ line with $(\pm\sqrt{-\beta/\eta},0)$. They satisfy the criterion $4\beta\eta<0$, manifesting a VHS at $E_{\xi}=0$ with a broadening [see Fig.~\ref{vanhove} (b)].  

We next discuss the critical case of three exceptional rings when two grey rings reside on both sides of the Fermi line, and the green ring coalesces on the Fermi line. In this situation, the saddle points critically get destroyed along the $k_x$ axis. Consequently, we obtain a single peak in the DOS, which at a higher resolution~\cite{yuan2019magic,sboychakov2022moire} eventually splits into two peaks with the emergence of higher order ($E^{1/3}$ and $E^{1/2}$) power law divergences [see Fig.~\ref{vanhove} (c)]. Finally, we consider the case where an exceptional ring consisting of two EPs of opposite chirality annihilate with the tuning of non-Hermiticity and light intensity. In this case, one grey ring crosses the Fermi line. Consequently, we obtain a saddle point at the high symmetry point $(0,0)$ with the condition $4 \delta_i \zeta_i<0$, resulting in a pair of VHS symmetric around the origin [see Fig. ~\ref{vanhove} (d)].

In conclusion, we have proposed an interface system of two NH nodal line semimetals, driven by light which enables control over higher-order VHS physics. In the Hermitian limit, the DOS shows VHS with logarithmic divergence. However, once non-Hermiticity is switched on, the two nodal rings produce four exceptional rings on two sides of the Fermi line. By tuning the light amplitude and non-Hermiticity, the occurrence of VHS is tuned with the motion and merging of the exceptional rings. We note here that both the tunable parameters, i.e., the light amplitude and the NH parameter are important for the unique behavior of the VHS in our analysis. The paired VHS, as well as higher-order VHS with $E^{1/3}$ and $E^{1/2}$ power-law divergences appear in the interface system depending on the values of the driving and NH parameters. Overall, our results suggest NH interfaces to be promising avenues for exploring higher-order VHS and their accompanying physics.

A promising avenue for experimentally realizing our proposal involves illuminating a junction between a topological insulator and a ferromagnet, which has been recently identified as a potential platform for realizing NH gapless phases by Bergholtz and Budich~\cite{bergholtz2019non}. Notably, this platform hosts EPs, and shining light on this junction presents a practical approach for implementing our proposed exceptional physics manipulation. Recent experimental demonstrations have employed angle-resolved scattering measurements, where, for instance, the emergence of a bulk Fermi arc connecting a pair of EPs was discovered by Zhou and colleagues~\cite{zhou2018observation}. Given the recent strides in experimental capabilities, we are hopeful that our predictions may be accessible in state-of-the-art platforms.

\noindent \textit{Acknowledgments:} A.B. is supported by the Prime Minister's Research Fellowship (PMRF). D.C. acknowledges financial support from DST (project number SR/WOS-A/PM-52/2019).  A.N. acknowledges support from a startup grant (SG/MHRD-19-0001) of the Indian Institute of Science.

\bibliography{references}

\end{document}